\documentclass[prd,preprintnumbers,floatfix,aps,notitlepage,nofootinbib,twocolumn,showpacs]{revtex4}

\usepackage{latexsym}
\usepackage{epsfig}
\usepackage{amssymb}

\newcommand{\lp}{\left(}
\newcommand{\rp}{\right)}
\newcommand{\lb}{\left[}
\newcommand{\rb}{\right]}

\newcommand{\ba}{\begin{eqnarray}}
\newcommand{\ea}{\end{eqnarray}}
\newcommand{\be}{\begin{equation}}
\newcommand{\ee}{\end{equation}}

\newcommand{\al}{\alpha}
\newcommand{\bt}{\beta}
\newcommand{\ga}{\gamma}

\newcommand{\la}{\lambda}

\newcommand{\La}{\Lambda}

\newcommand{\R}{\mathcal{R}}

\newcommand{\Q}{\mathcal{P}}
\newcommand{\ph}[1]{\phantom{#1}}

\begin{document}

\title{On new variational principles as alternatives to the Palatini method}

\author{Tomi Koivisto}
\email{T.S.Koivisto@uu.nl}
\affiliation{Institute for Theoretical Physics and Spinoza Institute, Leuvenlaan 4, 3584 CE Utrecht, The Netherlands.}
\date{\today}

\begin{abstract}

A variational principle was recently suggested by Goenner, where an independent metric generates the spacetime connection. It is pointed out here that the resulting theory is equivalent to the usual Palatini theory. However, a bimetric reformulation of the variational principle leads to theories which are physically distinct from both the metric and the metric-affine ones, even for the Einstein-Hilbert action. They are obtained at a decoupling limit of C-theories, which contain also other viable generalizations of the Palatini theories.

\end{abstract}

\maketitle

\section{Introduction}

Recently, an interesting new variational principle was proposed by Goenner as an alternative to the so called Palatini variational method \cite{Goenner:2010tr}. In the Palatini method, the metric and the connection are treated as independent variables, according to {\it the metric-affine variational principle}, in contrast to the usual metric variation where it is assumed a priori that the connection is the Levi-Civita connection computed from the metric \cite{Deser:2006ht}. In Goenner's variant of this principle, the connection is taken to be the Levi-Civita connection of another metric. Contrary to the claims made in Ref.\cite{Goenner:2010tr} however, both variants of the Palatini method yield an equivalent theory when applied to an $f(\R)$ action.

The concept of a connection as a fundamental variational degree of freedom but subjected to metric-compatibility appears to us somewhat vague. It remains unclear how to apply this restriction of variation in practice to a generic action, whilst the problem is absent in the $f(\R)$ case at the classical level since the extremals of the action turn out a posteriori to belong to a metric subspace. In any case, it is natural and technically straightforward to instead promote the underlying metric to a fundamental field with independent variations. For an obvious reason, we call this approach {\it the bimetric variational principle}. This indeed results in theories that are different from the corresponding metric as well as metric-affine ones, even in the case of Einstein-Hilbert action
where the two latter approaches are both well-known to lead to general relativity.

 It is illuminating to view this from the wider perspective of C-theories \cite{Amendola:2010bk}. This is a unified framework which includes as special cases all theories that emerge from the metric and the Palatini variational methods, and describes also Goenner's variant of the Palatini principle, or alternatively, its bimetric version, at a certain decoupling limit.
 As far as we know, Ref. \cite{Amendola:2010bk} was the first application of a bimetric variational principle, or ''formulation'' for short. Here we point out a subtlety which was missed there: whereas the metric-affine formulation of C-theories (trivially) reduces to the usual Palatini theory in the special decoupling limit, the bimetric formulation in general doesn't. The latter carries additional dynamics due to the second order property of the field equation. We also go further to generalize these results to actions beyond $f(\R)$.

\section{On metric-affine variations}

In order to clarify these issues, it is useful to first distinguish two different roles of a connection in gravitational theories. In the definition of the basic curvature object, the Riemann tensor, $\hat{R}^\alpha_{\beta\gamma\delta}$, one refers to a connection $\hat{\Gamma}$ as
\be \label{riemann}
\hat{R}^\alpha_{\beta\gamma\delta}=\hat{\Gamma}^\alpha_{\beta\gamma,\delta}-
\hat{\Gamma}^\alpha_{\beta\delta,\gamma}+
\hat{\Gamma}^\alpha_{\lambda\gamma}\hat{\Gamma}^\lambda_{\beta\delta} -
\hat{\Gamma}^\alpha_{\lambda\delta}\hat{\Gamma}^\lambda_{\beta\gamma}\,.
\ee
This tensor can be constructed in a metric-independent way by parallel transporting a vector $v$ around a closed curve, $[\hat{\nabla}_\mu,\hat{\nabla}_{\nu}]v^{\alpha}=\hat{R}^\alpha_{\gamma\mu\nu}v^\gamma$. The Riemann tensor is needed to compose invariants of the curvature. Thus, the connection $\hat{\Gamma}$ inevitably enters any covariant gravitational action and so affects the form of the left hand side of the gravitational field equations ensuing from that action. We therefore call $\hat{\Gamma}$ the {\it geometric connection}.

A crucial point is that, as long as there is a metric, another connection is always present. This is of course the Levi-Civita connection $\Gamma$ generated by the metric as
\be \label{matter_c}
\Gamma^\alpha_{\beta\gamma} = \frac{1}{2}g^{\alpha\la}\lp g_{\beta\la,\ga}+g_{\la\ga,\bt}-g_{\bt\ga,\la}\rp\,.
\ee
In the gravity theories we are discussing the metric is formally providing the invariant volume element and the contractions of vectors in the matter action. The metric thence provides measure of physical distances. Consequently, the trajectories of material particles are the shortest paths with respect to the metric. Equivalently, they follow the geodesics determined by the connection (\ref{matter_c}), which we hence call the {\it matter connection}. This result, implied by the so called generalized Bianchi identity that is a consequence of the diffeomorphism invariance of the action, is completely independent of the left hand side of the field equations, in particular the geometric connection that generates them \cite{Koivisto:2005yk}.

Using our notation, the starting point of Ref.\cite{Goenner:2010tr} may then be written as\footnote{In the notation of Ref.\cite{Goenner:2010tr}, $\hat{g}_{\mu\nu} \rightarrow g_{\mu\nu}$, $g_{\mu\nu} \rightarrow h_{\mu\nu}$ and $\hat{\Gamma} \rightarrow \{\}_g$. The two latter fields are the variational degrees of freedom in the approach of Ref.\cite{Goenner:2010tr}.}
\be \label{action}
S = \int d^4 x \sqrt{-g}\lb f(\R) + \mathcal{L}_m(\Psi,g_{\mu\nu})\rb\,,
\ee
where $\R=g^{\bt\delta}\hat{R}_{\bt\delta}$ and the tensor $\hat{R}_{\bt\delta}$ is understood to be a functional of an independent metric $\hat{g}_{\mu\nu}$. More precisely, it is given by the expression (\ref{riemann}) when $\alpha=\gamma$, and
\be \label{geometric_c}
\hat{\Gamma}^\alpha_{\beta\gamma} = \frac{1}{2}\hat{g}^{\alpha\la}\lp \hat{g}_{\beta\la,\ga}+\hat{g}_{\la\ga,\bt}-\hat{g}_{\bt\ga,\la}\rp\,.
\ee
Matter fields are collectively denoted by $\Psi$.  The significance of the metric $g_{\mu\nu}$ should be clear from the discussion above.
In fact, with respect to this metric, we have a metric theory of gravitation satisfying the three requirements that \cite{lrr-2006-3}
1) there exists a symmetric metric,
2) test bodies follow geodesics of the metric, and
3) in local Lorentz frames, the non-gravitational laws of physics are those of special relativity. On the other hand, then the independent metric $\hat{g}_{\mu\nu}$ may in a sense be regarded as an auxiliary field. It can be algebraically eliminated in favor of the observable metric $g_{\mu\nu}$, and it is most transparent to write the field equations in terms of the latter.

The opposite viewpoint is adopted in Ref.\cite{Goenner:2010tr}. The metric $g_{\mu\nu}$ is interpreted there as an auxiliary field devoid of physical significance and the field equations are written in terms of the other metric. Despite of being presented as a novel alternative to the previously considered gravity theories, the ensuing theory is nothing but the Palatini version of $f(\R)$  disguised in the Einstein frame. It was claimed that the equivalence principle is broken since matter stress tensor is not covariantly conserved with respect to $\hat{\Gamma}$, and that no matter gradients appear in the theory. The stress energy is however covariantly conserved with respect to $\Gamma$, and when written in terms of the metric that is compatible with this connection, the matter gradients appear explicitly into the theory. In this light, the misleading conclusions reached in Ref.\cite{Goenner:2010tr} stem from a confusion regarding the physical roles of the two connections, or in this case equivalently, the associated metrics\footnote{It is formally possible to identify the Einstein frame as the physical frame, and then the equivalence principle indeed appears to be violated. The Einstein frame version of Palatini-$f(\R)$ gravity (that is ''dynamically equivalent'' to the Jordan frame version) has been also considered in the literature \cite{Olmo:2011uz}. We are not concerned here with the choice of frame, which is a separate issue from the result that the two variational principles yield equivalent theories.}.

The rationale of C-theories, recently introduced by Amendola {\it et al} \cite{Amendola:2010bk}, is the possibility of a nontrivial relation between the matter and the geometric connections. As a simple example we first consider the following conformal relation: $\hat{g}_{\mu\nu}=C(\R)g_{\mu\nu}$. In particular, we look at a specific class of C-theory actions parameterized by $\al \in \mathbb{R}$ in such a way that when the parameter $\al=0$, the corresponding metric, and when $\al=1$, the corresponding Palatini theory is reproduced for an arbitrary function $f(\R)$. An example of such $\alpha$-parameterization is given by the exponential interpolating function as
\be \label{ctheory_c}
S_\alpha = S+\int d^4x\sqrt{-g}\la^{\mu\nu}_\rho\lp \hat{\Gamma}^\rho_{\mu\nu}-
\left\{ {\rho \atop \mu\nu} \right\}_{\lp f'(\R)\rp^\alpha g}\rp\,.
\ee
Thus, we add to the action (\ref{action}) a Lagrange multiplier which constraints the geometric  connection to be the Levi-Civita connection of the metric $\hat{g}_{\mu\nu}=\lp f'(\R)\rp^\alpha g_{\mu\nu}$. When $\al=0$, the two connections coincide and we have a metric theory.

When $\al=1$, the geometric connection is in the Einstein frame, which is the peculiar relation of Palatini-$f(\R)$ theories. However, in general also the lagrangian multiplier contributes to the dynamics. This can be of course changed by rescaling $\la^{\mu\nu}_\rho\rightarrow (1-\al)\la^{\mu\nu}_\rho$, and then we recover precisely Goenner's action (\ref{action}) at the limit $\al=1$.  This is a discontinuous limit of the theory, since a degree of freedom becomes nondynamical there. This decoupling is the culprit for the pathology of the $\omega=-3/2$ Brans-Dicke theory that was discovered decades ago and whose disturbing consequences have surfaced in many different contexts more recently \cite{Olmo:2011uz}.

The action (\ref{ctheory_c}), due to the presence of the lagrange multiplier may display no discontinuity in the propagating degrees of freedom in the limit $\al=1$. Still, the peculiar relation $C(\R)=f'(\R)$ holds and furthermore, $\la=0$ is always a consistent solution of this version of the theory. Thus the solutions of theory (\ref{action}) form a subset of the solutions of the new theory, which nevertheless seems to avoid the notorious theoretical and observational problems of the former. Hence the action (\ref{ctheory_c}) at $\alpha=1$ can realize the motivation of Ref. \cite{Goenner:2010tr} by introducing a phenomenologically viable alternative to the Palatini method.


\section{On bimetric variations}

As discussed in the introduction, one can also consider the set-up where the metric $\hat{g}_{\mu\nu}$ rather than the connection $\hat{\Gamma}$ is the fundamental field \cite{Amendola:2010bk}.
Instead of the action (\ref{ctheory_c}), where the gravitational degrees of freedom consisted of the triplet $({g}_{\mu\nu},\hat{\Gamma},\lambda)$, one would write the action
\be \label{ctheory}
S_\alpha = \int d^4x\sqrt{-g}\lb f(\R) + \la^{\mu\nu}\lp \hat{g}_{\mu\nu}-\lp f'(\R)\rp^\alpha g_{\mu\nu} \rp\rb\,,
\ee
the independent fields being now $({g}_{\mu\nu},\hat{\Gamma},\lambda)$. By erasing the lagrangian multiplier constraint one then obtains a bimetric reformulation of Goenner's starting point (\ref{action}).

One may also ask whether the conclusions persist beyond the $f(\R)$ theories. In particular, one could suspect that the conformal relation appearing in this special class of theories is necessary to guarantee the degeneracy of the two variational methods. Starting from more general forms of action, one obtains a more complicated relation between the independent and the metric connection by applying the metric-affine variational principle \cite{Olmo:2011uz}. In the remainder of this communication, we will generalize the bimetric variational principle and $C$-theory field equations to such actions.

For this purpose, we write the action
\be \label{action2}
S = \int d^4 x \sqrt{-g}\lb f(\R,\Q) + \mathcal{L}_m(\Psi,g_{\mu\nu})\rb\,,
\ee
allowing in there the invariant $\Q$ constructed from the two metrics via
\be
\Q = g^{\alpha\beta}g^{\gamma\delta}\hat{R}_{\al\ga}\hat{R}_{\bt\delta}\,,
\ee
where again $\hat{R}_{\mu\nu}=\hat{R}_{\mu\nu}[\hat{\Gamma}[\hat{g}]]$ when taking into account Eqs. (\ref{riemann}) and (\ref{geometric_c}).
The field equations for the metric $g_{\mu\nu}$ are
\be \label{field2}
\lp f_{,\R}\hat{R}_{\mu\nu}+2f_{,\Q}\hat{R}_{\mu\alpha}\hat{R}^{\nu\alpha} \rp - \frac{1}{2}f g_{\mu\nu} = T_{\mu\nu}\,,
\ee
where $T_{\mu\nu}$ is the matter stress energy tensor. The variation with respect to the metric $\hat{g}_{\mu\nu}$ yields
\be \label{field3}
\hat{D}^{\mu\nu}_ {\al\bt}\lb \sqrt{\frac{{g}}{\hat{g}}}\lp f_{,\R}g^{\al\bt} + 2f_{,\Q}\hat{R}^{\al\bt}\rp\rb = 0\,,
\ee
where we have defined the differential operator
\ba
\hat{D}^{\mu\nu}_{\al\bt}  =
\frac{1}{2}\sqrt{\frac{\hat{g}}{g}}\Big(\hat{g}^{\mu\nu}\delta^\rho_\alpha\delta^\gamma_\beta
& + & \hat{g}^{\rho\gamma}\delta^{\mu}_\alpha\delta^\nu_\beta \nonumber \\
  -   \hat{g}^{\rho\nu}\delta^{\mu}_\al\delta^{\gamma}_\bt & - & \hat{g}^{\rho\nu}\delta^{\mu}_\bt\delta^{\gamma}_\al
\Big) \hat{\nabla}_{\gamma}\hat{\nabla}_\rho\,.
\ea
The equation (\ref{field3}) is solved by
\be \label{disformal}
\hat{g}_{\mu\nu} = f_{,\R}g_{\mu\nu} + 2f_{,\Q}\hat{R}_{\mu\nu} \equiv h_{\mu\nu}\,.
\ee
This is the same disformal relation one obtains \cite{Olmo:2011uz} in the metric-affine variation of the action corresponding to (\ref{action2}).
Since, obviously in the usual Palatini version of the theory the field equations have the same the form (\ref{field2}), the dynamics of the solution (\ref{disformal}) coincide with those\footnote{
 We assume that in the metric-affine variation of the action (\ref{action2}) one is restricted to symmetric connections and Ricci tensors. Furthermore, in the bimetric variations discussed here, we assume the metric $\hat{g}_{\mu\nu}$ to be symmetric. Though, since this metric is related to the spin connection aspects of geometry and not to physical distances, it would be meaningful to relax this assumption, we omit exploring the possibility here.}.

There is however the following subtlety. Unlike from a metric-affine variation which yields first order equations of motion, we now obtained a second order equation which may have different solutions. Indeed, Eq.(\ref{field3}) appears to allow solutions where the nonmetricity of the connection $\hat{\Gamma}$ with respect to the metric $h_{\mu\nu}$ defined by (\ref{disformal}) is nonvanishing $Q_{\la\al\bt}=-\hat{\nabla}_\la h_{\al\bt} \neq 0$. By writing open eq.(\ref{field3}), one gets a nontrivial differential constraint on nonmetricity. In general then, the solutions of the theory need not coincide with the Palatini theory, either in its usual or in its C-theory form, since the connection need not be the Levi-Civita connection of $h_{\mu\nu}$. Even in the Einstein-Hilbert case, when $h_{\mu\nu}=g_{\mu\nu}$, there can be nonmetric solutions. They are classically distinguishable from general relativity (or from more general Palatini theories when $f\neq \R$) if the difference of the metrics doesn't correspond to a projective transformation of the geometric connection.
In general from Eq.(\ref{riemann}) one gets that
\be
\hat{R}_{\al\bt} = {R}_{\al\bt}(h)+\hat{\nabla}_\mu\Delta^\mu_{\al\bt}-\hat{\nabla}_\bt\Delta^\mu_{\mu\al}
+ \Delta^\la_{\la\rho}\Delta^\rho_{\al\bt}
- \Delta^\la_{\al\rho}\Delta^\rho_{\bt\la}\,.
\ee
where in the case at hand we have
\be
\Delta^\rho_{\al\bt} = \frac{1}{2}g^{\rho\la}\lp Q_{\al\bt\la}+Q_{\bt\al\la}-Q_{\la\al\bt}\rp\,.
\ee

Let us demonstrate explicitly the appearance of nonmetricity in the general relativistic Einstein-Hilbert action $f=\R$ within the bimetric formulation. For simplicity, we assume conformal nonmetricity described the function $b$ as $Q_{\la\al\bt}=-\hat{\nabla}_\la g_{\al\bt} = b_{,\la}g_{\al\bt}$. The field equations (\ref{field2}) and (\ref{field3}) for the two metrics $g_{\mu\nu}$ and $\hat{g}_{\mu\nu}$ become now, respectively
\be
G_{\mu\nu}+\frac{1}{2}b_{,\mu,}b_{,\nu}-\nabla_{\mu}b_{,\nu}+\lp \Box b+\frac{1}{4}(\partial b)^2\rp g_{\mu\nu} = T_{\mu\nu}\,, \label{1st}
\ee
\be \label{2nd}
 \nabla_\mu b_{,\nu} - 2b_{,\mu}b_{,\nu} -
\lp \Box b - \frac{1}{2}(\partial b)^2 \rp g_{\mu\nu} = 0\,.
\ee
There is no matter source in the right hand side of (\ref{2nd}) since we have assumed that matter is minimally coupled to geometry.
In fact by using the trace of the latter equation and rescaling $b=\sqrt{2/3}\phi$, one confirms the expectation that conformal nonmetricity contributes a canonic massless scalar obeying $\Box\phi=0$,
\be
G_{\mu\nu} - \phi_{,\mu,}\phi_{,\nu}+\frac{1}{2}(\partial \phi)^2 g_{\mu\nu} = T_{\mu\nu}\,. \label{1st_b}
\ee
Even without sources we can have propagating nonmetricity in the bimetric formulation of the action $f=\R$. In this sense the resulting theory is richer than in the metric or in the metric-affine formulation.
\begin{center}
\begin{table}[t]
\begin{tabular}{|c||c|c|c| }
\hline
 C-theory   & $\al \rightarrow 0$  & $\al \rightarrow 1$ & $\la \rightarrow 0$ \\
\hline \hline Metric-affine & metric & (1st order ?) gen. P. & Goenner $\sim$ P. \\
\hline Bimetric   & metric & (2nd order ?) gen. P. & New theories \\
\hline
\end{tabular}
\caption{A summary of our findings at the various limits of C-theories.
Metric-affine and bimetric formulations are considered. If we impose the metric constraint $\al=0$, the metric field equations follow in both cases.
We refer to theories where the $\hat{g}_{\mu\nu}=h_{\mu\nu}$ but there are more dynamics than in the usual Palatini theories as ''gen. P.''. In the decoupling limit $\la \rightarrow 0$ of the bimetric formalism, we may have solutions with $\hat{g}_{\mu\nu} \neq h_{\mu\nu}$. Such solutions are absent in the metric-affine formalism at this limit, where the theory reduces to the usual Palatini theory.  \label{tab}}
\end{table}
\end{center}
Finally, we look also at the C-theory generalization of the theory (\ref{action2}) in the form
\be \label{ctheory_g}
S  =  \int d^4 x \sqrt{-g}\lb f  + \hat{\lambda} - C\la - D\hat{\La} + \mathcal{L}_m(\Psi,g_{\mu\nu})\rb\,,
\ee
where for brevity of notation, we have suppressed the explicit dependence of the three functions $f$, $C$ and $D$ on the two scalars $\R$ and $\Q$, and introduced the invariants constructed by contractions of the field $\la^{\mu\nu}$ as
\be
\la = \la^{\mu\nu}g_{\mu\nu}\,, \quad
\hat{\la} = \la^{\mu\nu}\hat{g}_{\mu\nu}\,, \quad
\hat{\La} = \la^{\mu\nu}\hat{R}_{\mu\nu}\,.
\ee
Variations with respect to the three tensor fields $g^{\mu\nu}$, $\hat{g}_{\mu\nu}$ and $\la^{\mu\nu}$ give, respectively, the field equations
\ba
T_{\mu\nu} & = & L_\R \hat{R}_{\mu\nu} + L_\Q \hat{R}_{\mu\al}\hat{R}^{\al}_{\ph{\al}\nu}-\frac{1}{2}L g_{\mu\nu} + C\la_{\mu\nu}\,, \\
\la^{\mu\nu} & = & \hat{D}^{\mu\nu}_{\al\bt}\lb \sqrt{\frac{g}{\hat{g}}}\lp L_\R
g^{\al\bt}+2L_\Q\hat{R}^{\al\bt}-D\la^{\al\bt}
\rp\rb\,, \label{fe2} \\
\hat{g}_{\mu\nu} &= & Cg_{\mu\nu}+D\hat{R}_{\mu\nu}\,. \label{fe3}
\ea
We have employed the shorthand notations for the gravity lagrangian and its derivatives with respect to the two scalars
\ba
L &=& f  + \hat{\lambda} - C\la - D\hat{\La}\,, \\
L_\R & = & f_{,\R}  - C_{,\R}\la - D_{,\R}\hat{\La}\,, \\
L_\Q & = & f_{,\Q}  - C_{,\Q}\la - D_{,\Q}\hat{\La}\,.
\ea
Now the constraint (\ref{fe3}) dictates the relation of the metrics without any ambiguity. However, we should take into account that the lagrange multiplier has also other nontrivial consequences.
In the metric case $C=1$, $D=0$, $\la_{\mu\nu}$ contributes to the field equations in such a way that they reduce to the field equations one obtains from pure metric variation. In particular, $L=\R$ yields then pure Einstein gravity.

Choosing $C$ and $D$ suitably, we can obtain  $\hat{g}_{\mu\nu} = h_{\mu\nu}$, as in (\ref{disformal}).
 Also now, as in the metric-affine case above, the lagrangian multiplier can also contribute to the field equations.
 Its equation of motion is given by (\ref{fe2}). In the $f(\R)$ case, these theories can be described as biscalar tensor gravity, which do not in general reduce to the $\omega_{BD}=-3/2$ theory when $C=f'(\R)$.

We can thus straightforwardly construct C-theories which generalize (\ref{ctheory}) and interpolate between the modified Palatini theories and the metric theories. A possible choice is $C= f_{,\R}^\alpha$ and  $D= \alpha f_{,\Q}$.

The construction of the metric-affine C-theory generalizing (\ref{ctheory_c}) is completely analogous. There the rescaled theory $\lambda\rightarrow (1-\al)\la$ reduces to the usual Palatini theory at the decoupling limit $\al = 1$, but to recapitulate, in the bimetric framework this limit is different from both the usual Palatini and its C-theory generalisations.


\section{Conclusions}

We clarified the motivations and implications of recently introduced variational principles in gravity. Some erronous conclusions concerning (in)equivalences between theories were pointed out.

Starting from the distinction between the matter and the geometric connections, one may assume either the metric-affine or the bimetric formulation depending on whether the geometric connection or its underlying metric is the fundamental degree of freedom. We observed that in the latter case, even the Einstein-Hilbert action can support propagating torsion and nonmetricity, opening the possibility to observationally distinguish the correct fundamental assumption.

Furthermore, either formulation can be generalized by allowing an arbitrary relation between the two connections a'la C-theories. In particular, one can then obtain theories where the geometric connection is in the Einstein frame but which could avoid the worst problems of the usual Palatini theories. A summary of the different variants of the Palatini theory as special limits of C-theories is presented in the table \ref{tab}.

\bibliography{gammarefs}

\end{document}